\documentclass[11pt,a4]{interact}

\usepackage{epstopdf}
\usepackage[caption=false]{subfig}
\usepackage{caption} 
\usepackage{amsmath,amssymb,amsthm}
\usepackage{graphicx}
\usepackage{multicol}
\usepackage{flafter}
\usepackage{parskip}
\usepackage[ruled,vlined]{algorithm2e}
\usepackage{color,hyperref,xcolor}
\hypersetup{colorlinks=true,urlcolor=purple, citecolor=blue}
\usepackage{cleveref}
\usepackage{tablefootnote}

\usepackage{apacite}
\usepackage{natbib}
\newtheoremstyle{general}
{3mm} 
{3mm} 
{\it} 
{} 
{\bfseries} 
{.} 
{.5em} 
{} 

\theoremstyle{definition}

\theoremstyle{remark}

\numberwithin{equation}{section}

\def\P{\mathbb{P}}

\newcommand{\abs}[1]{\left\lvert #1 \right\rvert}
\newcommand{\norm}[1]{\left\| #1 \right\|}

\DeclareMathOperator*{\argmin}{arg\,min}

\begin{document}

\title{Optimal selection of the starting lineup for a football team}

\author{
\name{Soudeep Deb\thanks{CONTACT S. Deb. Email: soudeep@iimb.ac.in}, Shubhabrata Das\thanks{CONTACT S. Das. Email: shubho@iimb.ac.in}}
\affil{Indian Institute of Management Bangalore \\ Bannerghatta Main Road, Bangalore, 560076, Karnataka, India}}

\maketitle

\begin{abstract}
The success of a football team depends on various individual skills and performances of the selected players as well as how cohesively they perform. We propose a two-stage process for selecting optimal playing eleven of a football team from its pool of available players. In the first stage a LASSO-induced modified multinomial logistic regression model is derived to analyse the probabilities of the three possible outcomes. The model considers strengths of the players in the team as well as those of the opponent, home advantage, and also the effects of individual players and player combinations beyond the recorded performances of these players. In the second stage, a GRASP-type meta-heuristic is implemented for the team selection which maximises its probability of winning. The work is illustrated with English Premier League data  from 2008/09 to 2015/16. The application demonstrates that the model in the first stage furnishes valuable insights about the deciding factors for different teams whereas the optimisation steps can be effectively used to determine the best possible starting lineup under various circumstances. We propose a measure of efficiency in team selection by the team management and analyse the performance of the teams on this front.
\end{abstract}


%

\newpage

\section{Introduction}\label{sec:introduction}

Good cross-functional teams consist of members having different specialised skills. Success of such teams depends not only on the excellence of the individuals in their respective domains of speciality, but also on the teamwork. In this study, we focus on the problem of optimal team selection in the context of football (or, soccer). A football squad typically consists of a number of players, ranging from 20 to 30, who are specialists in different roles. Conventionally, there are four such positions -- goalkeepers, defenders, midfielders and forwards. For every match, the team management select a playing eleven (we use the term `starting lineup' interchangeably in this article) before the game starts. Our objective is to develop an optimisation algorithm that leverages historical data using sound statistical principle to provide the best possible starting lineup for a team depending on the opposition. 

Although our focus is on sports, we note that broadly similar problems have been addressed in literature in other aspects. \cite{Water2007} and \cite{Baykasoglu} proposed models for balanced team selection in the general context, with the latter using an algorithm based on fuzzy optimisation. \cite{Feng2010} used genetic algorithm for cross-functional team selection, while \cite{Alencar} dealt with selection of project team in multi-criteria decision making framework. 

In sports, a distinguishable feature is the binary or ternary outcome which results from the combined performance of the team. Despite having less ambiguity in terms of measuring the performance of a team from that perspective, the selection of the best possible lineup in such cases remains a challenging exercise. While many of the adopted approaches are intrinsic to specific sports under consideration, some of the adopted methodologies are more generic in nature. Data envelopment analysis (DEA) has been used by \cite{AMAGB2020} for player selection in cricket from the perspective of financial  performance in addition to on-field sports performances. \cite{Li2021} used DEA with multivariate logistic regression, and adopted a two-stage approach in predicting performance of basketball teams. \cite{Dadelo2014} and \cite{Toledano} also analysed basketball data, with the first employing a multi-criterion evaluation of players' and teams' performances; and the latter developing a performance index rating using multi-objective evolutionary algorithms. \cite{Bhattacharjee2015} and \cite{sharp2011integer} used integer programming for team selection in cricket. \cite{Budak2019} addressed the team selection problem in volleyball, and accommodated coaches' preferences while focusing on assigning roles to different players.  

In one of the earliest works related to football, \cite{Hirotsu} used dynamic programming in Markov processes to devise good strategy for formation based on starting line-up and substitutes. \cite{Ozceylan} used analytical hierarchical processing for football team formation, building  on the work of \cite{Boon2003EJOR} who used assignment based model in the general team building context. \cite{mchale2018identifying} used statistical model and network analysis on difficulty of pass to determine key players in a football team. \cite{Zeng} used optimisation for team composition in a  computer-based football game, pro-evolution-soccer. In a recent work, \cite{Zhou21}  adopted multi-criterion optimisation in football team selection using genetic algorithm under budget constraints. 

While the aforementioned papers have their own merits, there is a glaring drawback that the focus is generally on finding the best possible lineup only under different restrictions, and not on maximising the gain which is often associated with the most favourable outcome. To elaborate, it is often seen in football that a lineup consisting of the eleven best-rated players does not necessarily produce the highest chance of winning. A few studies, such as \cite{ge2020analysis}, \cite{cao2022team}, argue that in addition to player ratings, interaction and cooperation between players, opponent team, and other factors are also important. Extant literature on optimum lineup in football do not assess the effect of these additional variables on the most favourable outcome. We bridge this gap by constructing a methodology that finds deciding factors behind a team's success, and subsequently apply suitable techniques to deduce the optimal lineup that is most likely to win against a specific opponent. 

Our method is developed through a two-stage process. First, we arrive at a modified multinomial logistic regression (MLR) model that predicts the probability of win, draw and loss for a reference football team based on various skills and historical performances of the players, from both the reference team and the opposition. Standard extensions to MLR are infeasible in the present context because of a large number of potential regressors. To circumvent this, we combine the techniques of Least Absolute Shrinkage and Selection Operator (LASSO, proposed by \cite{tibshirani1996regression}) and the MLR to develop an appropriate statistical framework. Our prescribed modification to a standard MLR model enforces the choice of certain attributes that make potential difference to every selection or non-selection of players. By doing this, the method is able to identify interesting cases of a few individual players as well as some pairs of players whose inclusion in their teams' starting elevens affect the outcome probabilities significantly. In the second stage of our algorithm, we adopt a heuristic from the principles of discrete optimisation, where we follow a greedy random adaptive search procedure (GRASP) type algorithm (\cite{FEO1989}), with an aim to select the players in the reference team that maximise the probability of win against the specific opponent. Since the composition of the opponent is unknown to the reference team, we assume that they would select the most-skilled players for the respective formation. 

The basic idea of our heuristic stems from the proximity-optimality-principle, i.e.\ good solutions at one level or iteration is likely to be found `close to' good solutions at an adjacent level. While the optimal solution is not guaranteed in such iterations, random restarts of the search process can be effective.  Our multi-start heuristic starts with a random feasible solution which can be greedy in nature in terms of skills of the players. Then, we suitably define the neighbourhood of a solution such that only feasible solutions are considered. From a solution obtained at an iteration, only the most attractive neighbourhood solution is arrived at the next iteration. To avoid getting trapped in a local optimum, a new search loop with a fresh random solution is considered whenever the solutions converge in terms of objective function or the most attractive neighbourhood solution happens to be one among the last $k$ solutions (a user-defined criteria). While within a particular loop worse solutions are permitted, solution at any point of iteration refers to the best solutions encountered till that point. The algorithm stops based on specific convergence criteria related to the objective function, minimum number of random starts and minimum number of iterations.

The above methodology is explained in more detail in Sections \ref{subsec:model} and \ref{subsec:optimisation}. Before that, in \Cref{sec:materials}, we describe the data and notations used in this paper. The results and implications are discussed in \Cref{sec:results}. We conclude in \Cref{sec:conclusion} with important remarks on the significance of the results, possible modifications and extensions of the work.

\section{Data and notations}\label{sec:materials}

\subsection{Data description}\label{subsec:data}

The data used in the study is extracted from the European soccer database (ESD), publicly available in Kaggle. We use English Premier League (EPL) data from eight seasons (2008/09 to 2015/16). Every season, 20 teams participate in the EPL on a home-and-away basis. The bottom three teams are relegated to the lower division, while the top three teams of the lower division get promoted to the EPL for the next season. Keeping this in mind, we restrict to the ten teams which played in the top division in all of the aforementioned eight seasons. Although each such team played $38\times 8=304$ matches in this period, we only consider the matches for which complete information (about the teams and all players) were available in the database. These details, along with the summary of the results for the ten teams are presented in \Cref{tab:teamnames}. Note that the data until 2014/15 are used as the training set to develop the best model, and the last season is kept as test set. This provides 38 matches in the testing period for each team. We also explored other ways of train-test split, and the conclusions broadly remain the same. 

 \begin{table}[!ht]
 \centering
 \caption{Teams considered in this study, along with the corresponding data sizes and summary of the match results.}
 {\small
 \label{tab:teamnames}
 \begin{tabular}{l|cccc|cccc|cccc}
   \hline
   Team & \multicolumn{4}{|c}{All matches} & \multicolumn{4}{|c}{Home matches} & \multicolumn{4}{|c}{Away matches} \\
   & Total & Win & Draw & Loss & Total & Win & Draw & Loss & Total & Win & Draw & Loss \\  
   \hline
   Arsenal & 287 & 55.1\% & 25.1\% & 19.9\% & 144 & 63.2\% & 23.6\% & 13.2\% & 143 & 46.9\% & 26.6\% & 26.6\% \\ 
   Aston Villa & 272 & 25\% & 29.4\% & 45.6\% & 133 & 26.3\% & 32.3\% & 41.4\% & 139 & 23.7\% & 26.6\% & 49.6\% \\ 
   Chelsea & 288 & 56.6\% & 23.6\% & 19.8\% & 144 & 66\% & 22.2\% & 11.8\% & 144 & 47.2\% & 25\% & 27.8\% \\ 
   Everton & 290 & 39.3\% & 33.1\% & 27.6\% & 142 & 49.3\% & 28.9\% & 21.8\% & 148 & 29.7\% & 37.2\% & 33.1\% \\ 
   Liverpool & 286 & 47.9\% & 25.5\% & 26.6\% & 143 & 53.8\% & 30.1\% & 16.1\% & 143 & 42\% & 21\% & 37.1\% \\ 
   Man City & 277 & 57.4\% & 19.9\% & 22.7\% & 141 & 72.3\% & 12.8\% & 14.9\% & 136 & 41.9\% & 27.2\% & 30.9\% \\ 
   Man Utd & 285 & 61.4\% & 19.3\% & 19.3\% & 140 & 75\% & 10.7\% & 14.3\% & 145 & 48.3\% & 27.6\% & 24.1\% \\ 
   Stoke City & 277 & 32.5\% & 27.8\% & 39.7\% & 137 & 46\% & 27\% & 27\% & 140 & 19.3\% & 28.6\% & 52.1\% \\ 
   Sunderland & 275 & 27.3\% & 30.9\% & 41.8\% & 140 & 32.9\% & 32.9\% & 34.3\% & 135 & 21.5\% & 28.9\% & 49.6\% \\ 
   Tottenham & 283 & 49.1\% & 25.1\% & 25.8\% & 140 & 57.1\% & 24.3\% & 18.6\% & 143 & 41.3\% & 25.9\% & 32.9\% \\
    \hline
 \end{tabular}
 }
 \end{table}

For each match, the data provides the starting eleven and their positions. Albeit there can be finer positional assignment, we restrict to the primary four roles -- goalkeeper, defender, midfielder, forward. While this is forced by the limited information in the dataset, we emphasise that the proposed optimisation strategy can be easily extended for more detailed positional assignments too. Next, considering that our main focus is on identifying player efficiency with respect to the outcome of the game, we describe the skill variables (attributes) for all the players. 
Every player is assessed in 33 different aspects, and based on their real-life performances in various matches, these attributes are computed and dynamically updated in a scale of 1 to 99. We categorise the attributes into four groups, as presented below.

\begin{itemize}
    \item Goalkeeping: diving, handling, kicking, positioning, reflexes.
    \item Defensive: interceptions, marking, standing tackle, sliding tackle.
    \item Attacking: crossing, finishing, heading accuracy, volleys, curve, free-kick accuracy, shot-power, long-shots, penalties.
    \item General: short-passing, long-passing, ball-control, acceleration, sprint-speed, agility, reactions, balance, jumping, stamina, strength, aggression, positioning, vision, dribbling.
\end{itemize}

The values of the attributes are updated periodically to reflect if the players are  performing better or worse in particular aspects, potentially due to age, experience etc. We do not consider all the attributes separately, but summarise them in each of the above-mentioned categories. Specifically, for each goalkeeper, we calculate the weighted average of the goalkeeping attributes and call it his goalkeeping skill. For every other player, his defensive, attacking, and general skills are calculated by taking weighted averages of the respective attributes. The weights used in these calculations are obtained by applying principal component analysis (PCA) to the data of all the players. This is a common technique of finding the most appropriate linear combination of many variables, in terms of explaining the most variation, and is often used in similar sports data (e.g.\ \cite{ricotti2013analysis}). 
We consider the weights corresponding to the first principal component, and the values are reported in Table S2 of the supplement.

\subsection{Notations} \label{subsec:notation}

Let us denote the players in a team by $P_1,P_2, \ldots P_N$, and write
\begin{equation}
\label{eq:team-composition}
    {\cal N} = \{ P_1,P_2, \ldots P_N \} = {\cal G} \cup {\cal D} \cup {\cal M} \cup {\cal F}, 
\end{equation}
where ${\cal G}, {\cal D}, {\cal M}, {\cal F}$ are the subsets of goalkeepers, defenders, midfielders and forwards in the team. The sets ${\cal D}, {\cal M}$, and ${\cal F}$ are not necessarily disjoint as outfield players can be eligible to play in multiple roles. For example, a player who conventionally plays in the midfield is assumed to be eligible to play both as a defender and a midfielder if his defensive skill is found to be higher than his general and attacking skills. Thus, while we assume that a squad consists of $G$ goal-keepers, $D$ defenders, $M$ mid-fielders, and $F$ forwards, in general $G+D+M+F \geqslant N$, with equality holding only if no player is found to be eligible to play in more than one role.

Let $\bm v_1,\bm v_2, \hdots, \bm v_N$ be the respective valuations of the skill indices for the $N$ players. Note that each $\bm v_i$ is 4-dimensional, referring to the attributes listed above and aggregated as mentioned to yield the goalkeeping skill for a goalkeeper and defensive, attacking and general skills for all other players. We  write $\bm v_i=(v_{i1},v_{i2},v_{i3},v_{i4})'$, where the coordinates 1, 2, 3, 4 correspond to goalkeeping, defensive, attacking and general skills respectively. For ease of notation, these four dimensions are listed for all players, even if some (like goalkeeping skill of a forward) of them are not going to be used in the algorithm. 

A starting lineup (or a solution) $S$ is a subset (of cardinality 11) of ${\cal N}$, and is a union of disjoint subsets $\bar{G}, \bar{D}, \bar{M}, \bar{F}$, such that $\bar{G} \subseteq {\cal G}, \; \bar{D} \subseteq {\cal D}, \; \bar{M} \subseteq {\cal M}, \;  \bar{F} \subseteq {\cal F}$. The cardinality of $\bar{G}$ is always 1, while the cardinalities of $\bar{D}, \bar{M}, \bar{F}$ collectively determine the formation the team. Further note that a starting lineup $S$ can be represented by a binary $N$-tuple $( S[1], S[2], \hdots, S[N])$, with the $i^{th}$ element $S[i]$ being 1 if $P_i$ is included in that lineup, and being 0 otherwise. Accordingly, $S$ can be represented by $\{ P_i: S[i] = 1\}$, and the strength of the lineup is captured by the following set of 10 skill measures:
\begin{equation}
\label{eq:ten-skills}
    \begin{split}
        & \mathcal{X}_G(S) = \sum_{i: P_i \in {\bar{G}} } v_{i1} S[i] ; \\ 
        & \mathcal{X}_D^d(S) = \sum_{i: P_i \in {\bar{D}} } v_{i2} S[i], \; \mathcal{X}_D^a(S) = \sum_{i: P_i \in {\bar{D}} } v_{i3} S[i] , \; \mathcal{X}_D^g(S) = \sum_{i: P_i \in {\bar{D}} } v_{i4} S[i] ; \\
        & \mathcal{X}_M^d(S) = \sum_{i: P_i \in {\bar{M}} } v_{i2} S[i], \; \mathcal{X}_M^a(S) = \sum_{i: P_i \in {\bar{M}} } v_{i3} S[i], \; \mathcal{X}_M^g(S) = \sum_{i: P_i \in {\bar{M}} } v_{i4} S[i] ; \\ 
        & \mathcal{X}_F^d(S) = \sum_{i: P_i \in {\bar{F}} } v_{i2} S[i], \; \mathcal{X}_F^a(S) = \sum_{i: P_i \in {\bar{F}} } v_{i3} S[i], \; \mathcal{X}_F^g(S) = \sum_{i: P_i \in {\bar{F}} } v_{i4} S[i].
    \end{split}
\end{equation}

Observe that the above terms represent in sequence the goalkeeping skill of the goalkeeper, average defensive, attacking and general skills of the defenders, those for the midfielders, and the same for the forwards. We believe that these consolidated measures provide a more consistent and useful evaluation of the performances instead of the original set of 33 skill variables.

\section{Methodology}\label{sec:methods}

\subsection{Stage 1: Model development}\label{subsec:model} 

Football is a team sport where individual brilliance in performance is expected to have a positive impact, but does not guarantee a positive result for the team. Keeping this in mind, we aim to find a suitable multinomial logistic regression (MLR) model that can predict the probability of win, loss and draw for the reference team. To that end, the first step is feature selection. 

We use $Y_i$ to denote the outcome (win, draw or loss) of the $i^{th}$ match in the dataset for a reference team. Many researchers (see, e.g.\ \cite{pollard1986home} and \cite{nevill1996factors}) established that playing at home almost always increases the chances of winning for a team. That motivates us to include a binary variable as a regressor, indicating whether it was home or away game for the reference team. The ten skill measures defined in (\ref{eq:ten-skills}), for both the reference team and the opposition are included in the feature set. Additionally, to account for the individual impact of the players in different positions, we take a fixed effects approach. To avoid identifiability and singularity issues, these fixed effects are considered only for the players who take corresponding positions in at least 30 matches. This allows a player to have varying impacts on the results based on what position he plays in a match. It is believed that in a team sports like football, some players can perform better when they play alongside specific teammates whereas their performances significantly deteriorate if they have to play without those teammates. This phenomena stems from the fact that the players often develop great understanding by practising and playing together over a long period of time. It is in fact closely connected with the psychological concept of team synergy. The reader is referred to \cite{araujo2016team} for in-depth discussions on team synergy in sports. With this hypothesis in view, we also include two-way interaction effects of the players. Akin to before, here also we consider the pairs who appear together in at least 30 matches for the reference team. The choice of 30 as the minimum number of matches required to be eligible for consideration of additional effects is subjective, and can be modified. 

Let  $m$ denote the total number of features (i.e.\ home factor, strength measures for the reference team as well as the opponent, fixed effects and two-way interaction effects of the players) corresponding to the model for one particular team. We  denote the vector of covariates for the $i^{th}$ match ($1\leqslant i \leqslant n$) by $\bm{x}_i=(x_{i1},x_{i2},\hdots,x_{im})^\top$, and the overall design matrix is given by $\bm{X}=[\bm{x}_1:\hdots:\bm{x}_n]^\top$. The vector of the outcomes is $\bm Y=(Y_1,\hdots,Y_n)^\top$. To follow the framework of MLR, taking `Draw' as the pivot category, we can write 
\begin{equation}
    \label{eq:MLR-model}
    \log \frac{P(Y_i=\text{Win} \mid \bm{x}_i)}{P(Y_i=\text{Draw} \mid \bm{x}_i)} = \bm{x}_i^\top\bm\beta_w, \; \log \frac{P(Y_i=\text{Loss} \mid \bm{x}_i)}{P(Y_i=\text{Draw} \mid \bm{x}_i)} = \bm{x}_i^\top\bm\beta_l,
\end{equation}
which can be restructured as
\begin{equation}
\label{eq:MLR-model-prob}
\begin{split}
    &P(Y_i=\text{Win}) = \frac{\exp(\bm{x}_i^\top\bm\beta_w)}{1+\exp(\bm{x}_i^\top\bm\beta_w)+\exp(\bm{x}_i^\top\bm\beta_l)}, \\ 
    &P(Y_i=\text{Loss}) = \frac{\exp(\bm{x}_i^\top\bm\beta_l)}{1+\exp(\bm{x}_i^\top\bm\beta_w)+\exp(\bm{x}_i^\top\bm\beta_l)}.
\end{split}
\end{equation}

The regression coefficients $\bm{\beta}_w$ and $\bm\beta_l$ are both $m$-dimensional, and $\bm\beta = (\bm{\beta}_w^\top, \bm{\beta}_l^\top)^\top $ represents the vector of all parameters of interest. Note that $m$ is  substantially bigger than $n$, thereby making it an infeasible problem to solve through the classical way. Thus, we adopt the LASSO approach as in a high-dimensional setting, LASSO helps in selecting appropriate features by shrinking the unimportant parameters to zero. We tweak the method 
by imposing the additional constraint that the goalkeeping skill of the goalkeeper, defensive skill of the defenders, general skill of the midfielders and the attacking skill of the forwards, both for the reference team and for the opposition, must  be selected in the model. This constraint is not only intuitively appealing, but it also ensures that selection and non-selection of every player has potential impact on the probability of each outcome. Further, we select the optimum regularisation parameter with the restriction that at most 20 variables can be selected in the best model. In this manner, the most important variables can be identified through a scientific way, while maintaining that the model complexity is 
tractable from an implementation perspective. 

\cite{friedman2010regularization} detailed the implementation of LASSO in a generalised linear model framework, and we follow their recommendations. We omit the finer details, except pointing out that LASSO, along with our constraints, works around the minimisation problem
\begin{equation}
\label{eq:normeq}
  \hat{\bm\beta} = \argmin_{\bm\beta} \{ - \log L(\bm{\beta})+ \lambda\norm{\bm\beta_S}_1\},
\end{equation}
where $L(\bm{\beta})$ is the complete likelihood for the multinomial data, $\bm\beta_S$ is the set of coefficients corresponding to the features on which we do not put any constraint, and $\norm{\cdot}_1$ is the $\mathcal{L}_1$-norm, i.e.\ the sum of the absolute values of the coordinates. 
We may note that the nonzero estimates from the LASSO approach have a tendency to be biased towards zero (\cite{meinshausen2007relaxed}). Thus, although it takes care of a large number of regressors, it warrants further improvement through an appropriate method of debiasing. With that in view, after implementing the LASSO step to select the variables, we fit the aforementioned MLR model on the data, which helps us in computing the probability of win, loss and draw depending on the values of the regressors.

\subsection{Stage 2: optimisation algorithm}\label{subsec:optimisation} 

Our primary goal in developing the optimisation algorithm is to maximise the probability of the reference team winning. Obviously, the method can be easily extended for other objectives, which we explicate in \Cref{sec:conclusion}. In the following, we consider optimal selection of a starting eleven for a given formation, typically expressed as $d-m-f$, where the three numbers stand for the numbers of defenders, midfielders and forwards, respectively. Note that every team management tend to play in a specific formation in most of the matches. However, if the interest is in the overall optimal solution, then it would suffice to search among the optimal solutions corresponding to the eight conventional formations (see below) or the appropriate subset of formations as the case may be. An all-exhaustive search among all possible squads is not computationally feasible because the number of such squads is too many. For illustration, consider a 25-person-squad with 3 goalkeepers, 8 defenders, 8 midfielders and 6 forwards. Here, even with the assumption of unique role for each player, the number of possible lineups under conventional formations 3-4-3, 3-5-2, 4-3-3, 4-4-2, 4-5-1, 5-2-3, 5-3-2, and 5-4-1  are 235200, 141120, 235200, 220500, 70560, 94080, 141120 and 70560, respectively. Clearly, a good optimisation strategy is necessary. 


In our methodology, for additional flexibility, we consider that a player (other than goalkeeper) can play in multiple positions. Recall that for outfield players, we compute the defensive skill, general skill and the attacking skill. Based on these variables, we consider that any player can be slotted in a defending role even if he conventionally does not play as a defender, if his defensive skill is more than general or attacking skills. Similarly, players with higher general skills are assumed to be suitable in a midfield role while higher attacking skills indicate a possible forward role for any player, in addition to his original role. In our algorithm, we take this aspect of a player's suitability in multiple roles into account and consider such possibilities while optimising over possible lineups.

Heuristic of the proposed algorithm has been outlined in \Cref{sec:introduction}. We reiterate that in the algorithm, feasible solutions are obtained for a number of iterations. As stopping rules, we set an apt convergence criterion as well as a minimum number of iterations. We also insist that a minimum number of random restarts are implemented to avoid being trapped into a local optimum. In our computation, the minimum number of iterations is set to 20 while we require a minimum number of 10 random starts. On the other hand, the main convergence criterion used in the algorithm is based on the functional value of the probability of win, denoted as $\P(win \mid S)$ for a feasible solution $S$. This probability is calculated through the LASSO-induced MLR model, as discussed in the previous subsection. 

At the first three iterations of the optimisation algorithm, candidate solutions $S_1$, $S_2$ and $S_3$ are selected randomly from feasible solutions. This is done by selecting requisite number of players from each of the four categories by simple random sampling with replacement. For the subsequent iterations, only the (feasible) neighbours of $S_i$ are considered and the best among them are adopted as $S_{i+1}$, unless it happens to be one among the previous solutions or the convergence criteria suggests that the algorithm has converged to a local optimum. Note that the first step is used to avoid being trapped in a loop. In either case, the algorithm then calls for a random restart and it is continued till a minimum number of random solutions are traversed. 

The proposed meta-heuristic is critically dependent on suitable definition of neighbours. Formally,  two solutions (or, starting lineups) $S$ and $S^*$  are defined as neighbours if they differ by only a single player, i.e.\
\begin{equation}
\label{eq:neighbour}
    \abs{S - S^*} = \sum_{i=1}^N \abs{S[i] - S^*[i]} = 2. 
\end{equation}  

Note that while implementing the optimisation algorithm, we let the win probability to be worse at the successive iteration to avoid getting trapped into  possible locally optimum solutions. It does not affect the eventual solution since at any stage the algorithm considers the best solution up to any iteration as its candidate solution. For better understanding of the reader, the entire pseudo-code of the two-stage algorithm is presented in Algorithm \ref{algo:optimal_squad}.

\begin{algorithm}[!ht]
\caption{Find the optimal starting lineup for a football team.}
\label{algo:optimal_squad}
\addtolength{\algomargin}{\parindent}
\SetKwFor{For}{for}{do}{end for}%
\SetKwFunction{cusum}{CUSUM}
\SetKwProg{Fn}{Function}{:}{}
\DontPrintSemicolon
\SetAlgoLined
\SetKwInOut{Data}{Input}
\Data{Training data on match results (response variable), along with all possible regressors (match details, aggregated player strengths, individual player effects and interaction effects). Detailed information on the next match (test set) are also provided.} 
\SetKwInOut{Result}{Output}
\Result{Fitted model and optimal starting eleven that maximises the win probability for the next match.}
    \textbf{Stage 1: Model development}: 
    \begin{enumerate}
    \itemsep0em 
        \item \label{st1} Fix the set of covariates $X$, including strength variables $F$ (details mentioned in \Cref{subsec:model}) that must be included in the model, dummy variable representing home-game or otherwise for the reference team, fixed effects of all qualifying player-position combinations, as well as their interaction effects. 
        \item \label{st2} Run LASSO in a generalised linear model framework to select the appropriate set of variables $\mathcal{V}$, by imposing the additional constraints that $\mathcal{V}$ must be a superset of $F$ and that the total number of selected variables must not be more than a desired number.  
        \item \label{st3} Run multinomial logistic regression model for the response variable with $\mathcal{V}$ as the feature set that would compute  $\P\left(win \mid S \right)$ 
        for any feasible solution $S$.
    \end{enumerate}

    \textbf{Stage 2: Run the optimisation algorithm}: 
    \begin{enumerate}
    \setcounter{enumi}{3}
    \itemsep0em 
        \item \label{st4} Set stopping criteria: 
    \begin{itemize}
    \itemsep0em 
        \item min\_it ($i_0$): Minimum number of iterations;
        \item prob\_diff ($\delta$): Minimum acceptable bound of relative improvement in the probability of win between consecutive iterations;
        \item min\_random\_start ($r_m$): Minimum number of random starts to the algorithm. 
    \end{itemize}
    
     \item \label{st5} Initiate iteration number $i=1$,  random\_starts $r=0$, flag $f=0$ .
        \item  \label{st6} Randomly select a feasible solution (starting eleven) and call it $S_i$; set $r=r+1$.
        \item \label{st7} Find $\P(win \mid S_i)$ using step \ref{st3}. 
        \item   \label{st8} Set $i=i+1$.  
        \item \label{st9} if $ i > 3 $, go to step \ref{st10}, else go to step   \ref{st6}.
        
        \item \label{st10} Find the list of neighbours of $S_{i-1}$ and find out the neighbour for which the probability of win is maximum (in case of tie, select one neighbour randomly). Call it $N(S_{i-1})$. 
        \item \label{st11} If $N(S_{i-1})$ is different from  ($S_{i-1}$ and) $S_{i-2}$ and $S_{i-3}$, and if 
        $$\frac{\abs{\P\left(win \mid S_{i-1}\right) - \P\left(win \mid N(S_{i-1}) \right)}}{\P\left(win \mid S_{i-1}\right) } > \delta,$$
        set $S_{i}=N(S_{i-1})$ and flag $f=1$.
        
        Else, randomly select a feasible solution and assign it to $S_{i}$. Find $\P(win \mid S_i)$ using step \ref{st3}. Set $r=r+1$, $f=0$. 
        
        \item \label{st12} Set $i=i+1$. Repeat steps \ref{st10} and \ref{st11} if $f=1$, or $r < r_m$, or $i < i_0$. Else, move to step \ref{st13}.
        
        \item \label{st13} Find $S_j$ such that $\P\left(win \mid S_j\right)$ is maximum for all $1 \leqslant j < i$. Return $S_j$ as the output.
    \end{enumerate}
\end{algorithm}

\section{Results}\label{sec:results} 

\subsection{Summary of the models}\label{subsec:model_summary}

We start with a brief summary of the details of the models estimated through our approach. As is well known, the coefficients and the significance of attributes in a regression setup depends on the presence or absence of other explanatory variables in that model. Consequently, suitable caution should be exercised while commenting on the presence, absence or significance of the attributes in the eventual models, even if they are not explicitly highlighted in our discussions. Moreover, to complement the analysis in this section and to show the efficacy of the proposed model, we conduct a comparative study with benchmark techniques. In the interest of space, these discussions are deferred to Section S.1 of the supplement. We want to point out that the proposed model records superior performance than the other techniques. The results also establish that a single model for all teams taken together is found to be less effective than developing separate models for each team.

Our data analysis steps, as described in the previous section, created 600 possible player-specific fixed effect variables and 2171 interaction effect variables in total. \Cref{tab:model_summary} shows the number of features selected by the LASSO step, along with the number of individual player-wise fixed effects and the number of two-way interaction effects (pairwise effects of the players), for all the teams. In the same table, we also present the values of the Akaike information criterion (AIC) and the estimated home effects (when the variable is selected in the model).

 \begin{table}[!htb]
 \centering
 \caption{Overview of the models for all the teams.} 
 \label{tab:model_summary}
 \begin{tabular}{lccccc}
   \hline
   Team & \multicolumn{3}{c}{Number of features}  & AIC & Home effect \\  
    & Selected & Individual & Pairwise & & Win / Loss \\ 
   \hline
   Arsenal & 12 & 0 & 2 & 475.69 & $0.42 (0.331) $ / $-0.75 (0.424) $ \\ 
   Aston Villa & 14 & 1 & 3 & 499.09 &  \\ 
   Chelsea & 12 & 1 & 2 & 447.94 & $0.74 (0.355) ^*$ / $-0.91 (0.467) $ \\ 
   Everton & 14 & 0 & 4 & 541.28 & $0.86 (0.326) ^*$ / $-0.48 (0.375) $ \\ 
   Liverpool & 12 & 1 & 1 & 504.18 & $-0.02 (0.332) $ / $-1.19 (0.404) ^*$ \\ 
   Man City & 12 & 1 & 2 & 436.29 &  \\ 
   Man Utd & 18 & 0 & 9 & 417.94 & $2.03 (0.438) ^*$ / $0.71 (0.548) $ \\ 
   Stoke City & 20 & 2 & 8 & 469.05 & $1.1 (0.396) ^*$ / $-1.11 (0.411) ^*$ \\ 
   Sunderland & 15 & 3 & 3 & 518.76 & $0.21 (0.368) $ / $-0.69 (0.338) ^*$ \\ 
   Tottenham & 15 & 0 & 5 & 485.83 & $0.45 (0.364) $ / $-0.63 (0.417) $ \\ 
    \hline
 \end{tabular}
 \end{table}

The models for Arsenal, Chelsea, Liverpool and Manchester City include the least number of variables. For Arsenal and Liverpool, only two player-specific features (including individual and pairwise) are important. In contrast, the outcome probabilities for Manchester United, Stoke City and Sunderland are found to be affected by several player-specific features. An interesting observation is that several players played in more than one position for enough games, but none had differential impact on the outcome while playing in multiple roles. Perhaps this indicates that no player can be crucial for their team in multiple roles. In this light, our model can suggest the most suitable role for a player in a team.

As many as nine interaction effects are relevant for Manchester United. In general, for all cases, general ratings and the pairwise features are more crucial than the individual fixed effects. For conciseness of the paper, detailed results on the estimates of all the coefficients, their standard errors and significance are deferred to the supplement, in Tables S3 to S12. Below, we focus on some of the interesting aspects based on the significance of these coefficients.

First, reflecting at the home advantage, it is interesting to observe that contrary to common perception and overall differential proportion of win/draw results as captured in \Cref{tab:teamnames}, home factor is found to be significant only for six teams. In case of Aston Villa and Manchester City, this variable is not selected by the proposed methodology. It suggests that once the effects of the players from the both sides are taken into consideration, the home advantage is not as important a factor as it is typically perceived. For the eight teams where the home factor is included in the model, the signs of the coefficients for this regressor are as expected, i.e.\ playing at home typically should improve the chance of win and reduce the chance of loss. We see that the home factor is most significant for Manchester United, followed by Stoke City, Everton and Chelsea. We also find that Liverpool and Sunderland have a significantly higher chance of losing away, although the results do not display a decided advantage at home.

\begin{figure}[!htb]
     \centering
     \caption{Effects of the forced variables in the models for different teams. Rows correspond to the goalkeeper strength, defenders' defensive strength, midfielders' general strength, and forwards' attacking strength. Effects on win/loss probabilities are displayed by different colours, the effects of own team's strength versus opposition's strength are presented in the two columns.}
     \includegraphics[width=16cm,keepaspectratio]{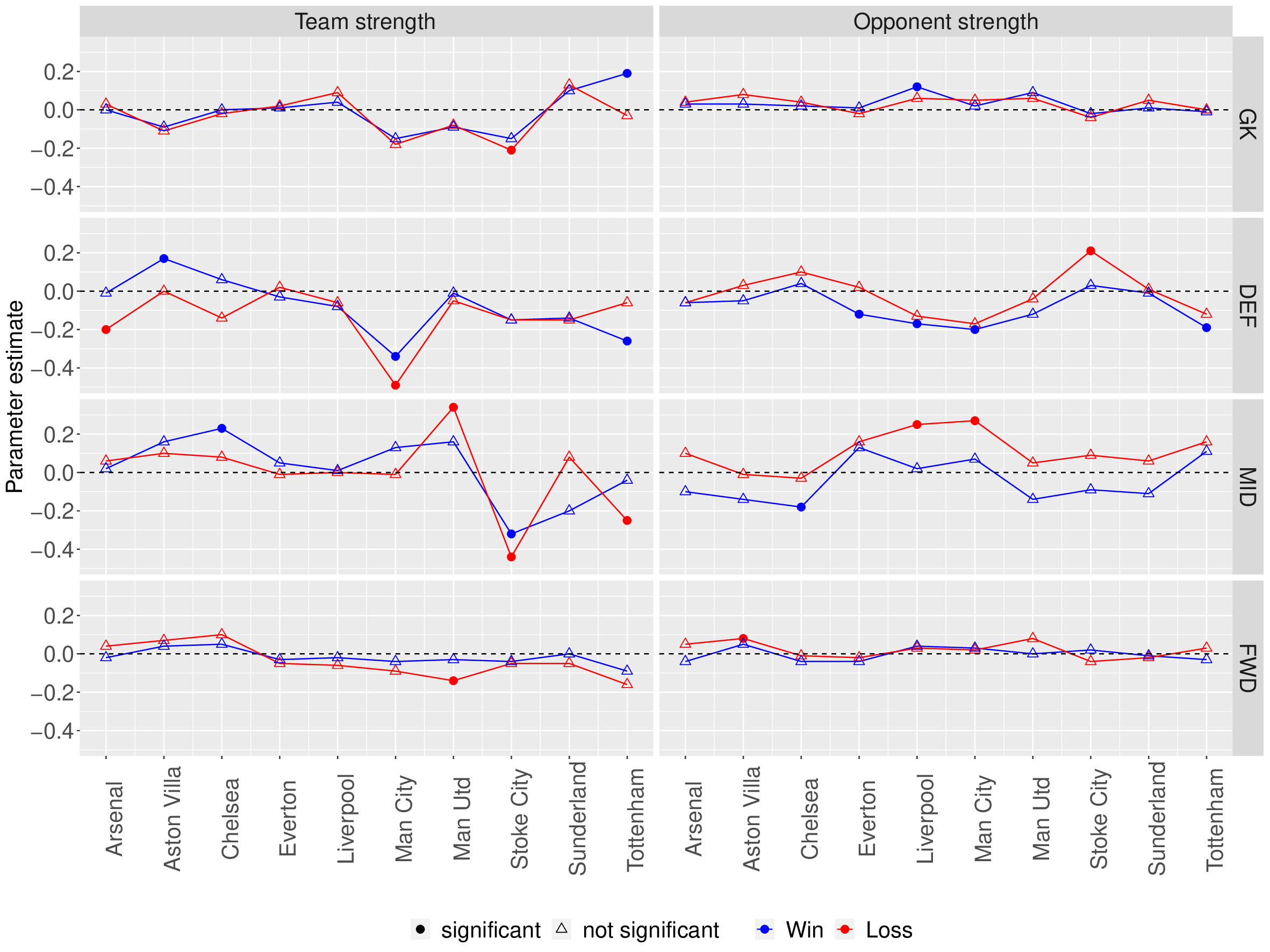}
     \label{fig:forced_var_plot}
 \end{figure}

Next, we turn attention to the coefficients of the eight variables that determine the strengths of the teams in the four different positions. Recall that inclusion of these variables has been made mandatory in order to guarantee that the selection or omission of every player affects the probability of the match outcomes. The coefficient estimates of these variables for all teams are presented in \Cref{fig:forced_var_plot}. There, blue and red colours indicate the effects on the probabilities of win and loss respectively.

Focusing on the bottom panel of the figure, we notice that the attacking strengths of forwards do not impact any team significantly, except for Manchester United and Aston Villa. For Manchester United, a better attacking lineup decreases the loss probability, while for Villa, the loss probability increases if the opponent forwards have higher attacking skills. Looking at the goalkeeping strength on the left side, we see that it is insignificant for all teams barring the case of Tottenham and Stoke City. Interestingly, we find that Tottenham can increase its win probability whereas Stoke City can decrease its loss probability with a better goalkeeper. In terms of the opposition strengths in defense and midfield, more teams are found to be impacted. It is natural that the significant effects of better defensive skills of the opponents are positive on the loss probability and negative on the win probability. Akin to this, we see that Arsenal's loss probability decreases significantly with lower defensive skills of the defenders, whereas a better midfield lineup in terms of its general skill improves Aston Villa's win probability and decreases Tottenham's loss probability significantly.

We also note a couple of counter-intuitive observations in case of Manchester United and Tottenham. For the former, higher general skill of the midfielders makes a significant positive impact on the loss probability, while for the latter, better defensive strength of defenders is identified to be a negative factor for the win probability. Finally, from the tables given in the supplement, one may notice that non-mandatory skill variables have been selected for Arsenal, Aston Villa, Everton, Liverpool, Manchester City, Stoke City and Tottenham. Most common such variables are the defensive strengths of the midfielders and the same of the attackers. This can be interpreted as better defensive abilities from the players across positions being instrumental in getting favourable results for some teams.

Moving on to the fixed effects and the interaction effects of the players, for brevity, in \Cref{tab:sig_pleff}, only the players and the pairs who have significant effects are listed. All coefficients are provided in the supplemental tables.

 \begin{table}[!htb]
 \centering
 \caption{Significant effects of individual players and combinations of players, as identified by the models. Standard errors are given in parentheses, * indicates significance at 5\% level.} 
 \label{tab:sig_pleff}
 \begin{tabular}{llcc}  
     \hline
     Team & Players (positions) & Win & Loss \\
     \hline
     Chelsea & Florent Malouda(M) & $-1.33 (0.55) ^*$ & $-1.27 (0.662) $ \\ 
             & Ashley Cole(D):Juan Mata(M) & $1.67 (0.677) ^*$ & $1.03 (0.802) $ \\ 
             & Cesar Azpilicueta(D):Ramires(M) & $-0.39 (0.431) $ & $-1.66 (0.648) ^*$ \\ 
     \hline
     Everton & Phil Jagielka(D):Sylvain Distin(D) & $-0.46 (0.354) $ & $-0.79 (0.399) ^*$ \\ 
             & Tim Howard(G):Steven Pienaar(M) & $-0.87 (0.372) ^*$ & $-0.89 (0.411) ^*$ \\ 
     \hline
     Liverpool & Albert Riera(M) & $-1.29 (0.544) ^*$ & $-3 (0.917) ^*$ \\ 
             & Fernando Torres(F):Dirk Kuyt(M) & $1.85 (0.674) ^*$ & $2.37 (0.762) ^*$ \\
     \hline
     Man City &   Shaun Wright-Phillips(M) & $-1.24 (0.654) $ & $-2.26 (0.843) ^*$ \\ 
     \hline
     Sunderland & Mamady Sidibe(F) & $-2.24 (0.779) ^*$ & $-4.03 (0.839) ^*$ \\ 
                & Robert Huth(D):Marc Wilson(D) & $-0.8 (0.562) $ & $-1.51 (0.553) ^*$ \\ 
               & Jonathan Walters(F):Glenn Whelan(M) & $0.15 (0.503) $ & $-2.34 (0.616) ^*$ \\ 
               & Ricardo Fuller(F):Matthew Etherington(M) & $-0.56 (0.658) $ & $-2.41 (0.786) ^*$ \\ 
               & Glenn Whelan(M):Rory Delap(M) & $-0.15 (0.577) $ & $2.01 (0.63) ^*$ \\ 
     \hline
     Stoke City & Jack Colback(M) & $1.8 (0.662) ^*$ & $0.17 (0.68) $ \\ 
     \hline
   Tottenham & Younes Kaboul(D):Aaron Lennon(M) & $1.36 (0.63) ^*$ & $1.73 (0.668) ^*$ \\ 
         & Ledley King(D):Benoit Assou-Ekotto(D) & $1.86 (0.574) ^*$ & $1.48 (0.627) ^*$ \\ 
     \hline
 \end{tabular}
 \end{table}

For Arsenal and Aston Villa, although several player-specific effects are selected by the model, they are not found to have a significant effect. In case of Chelsea though, all such variables are significant. The presence of Florent Malouda has an adverse impact on his team's probability of win, whereas the partnership of Ashley Cole - Juan Mata and the combination of Cesar Azpilicueta - Ramires have worked well for Chelsea. It makes sense since both pairs typically play in the same side, one behind the other. The defensive combination of Phil Jagielka and Sylvain Distin significantly reduces the loss probability for Everton. The same phenomena is observed in the partnership of Sunderland's Robert Huth and Marc Wilson too. Similarly, the midfielder-forward pairs of Matthew Etherington - Ricardo Fuller and Glenn Whelan - Jonathan Walters impact the outcomes in the same way. In contrast, Whelan, when paired with Rory Delap in the midfield, increases the loss probability significantly. There are two midfielders -- Shaun Wright-Phillips and Jack Colback -- who help their respective teams' causes in significant way when they are included in the starting lineups. Albert Riera is another midfielder who has significant impact on the outcome of Liverpool matches, and the coefficients are negative for both win and loss. By looking at the values of the coefficients, we may further say that the effect on reducing the  probability of loss is more prominent than the probability of win for this player. Analogous conclusion can be drawn about Sunderland's Mamady Sidibe.

\subsection{Results of the optimisation algorithm}
\label{subsec:results_optimisation}

In this subsection, we explore the outputs of the proposed optimisation algorithm in various aspects. Recall that we use the data from seven seasons to train the model. Then, for each match in 2015/16 season, the algorithm is implemented in a recursive manner to find out the optimal team that would maximise the win probability, assuming that the opposition would field their best eleven players. We emphasise that the algorithm can be easily used under alternative settings, e.g,\ assuming that the opposition chooses the lineup from the last match or the eleven most frequently used players. 

\begin{figure}[!htb]
     \caption{Comparison of the win probability (against best-rated opposition) for the optimal team with the same for the originally selected team, for each club and for all matches in the 2015/16 EPL season. Bookmakers' odds are shown with dotted lines.}
     \label{fig:2lineplot_prob}
     \centering
     \includegraphics[width = 13cm,keepaspectratio]{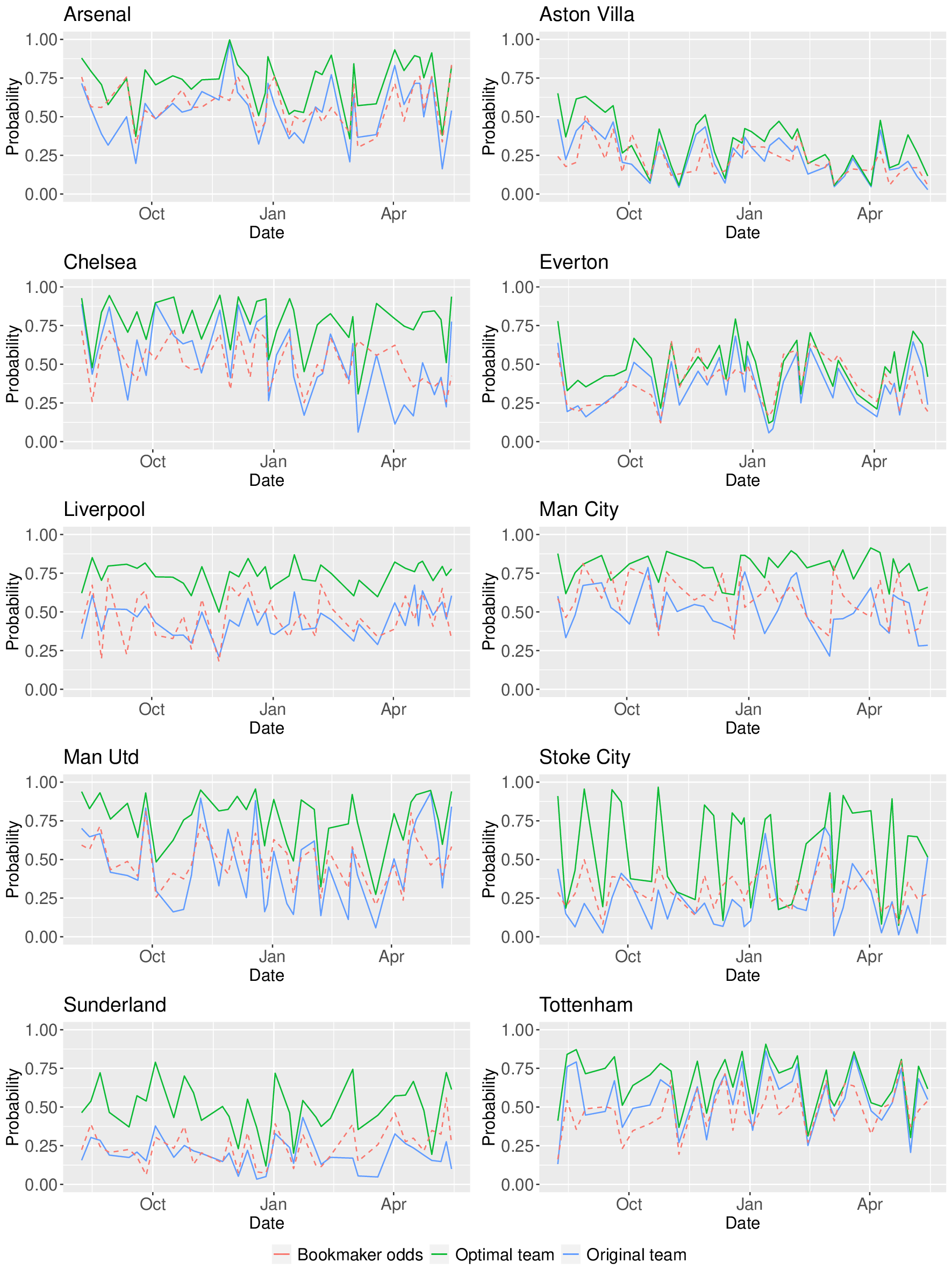}
 \end{figure}

In \Cref{fig:2lineplot_prob}, we present the probability of win for every team in the matches in the test set corresponding to the original starting eleven, and show the comparison against the maximum win probability possible by those teams, as calculated by the proposed algorithm. Additionally, the bookmakers' probabilities for the teams are also displayed in the same graph. 
As expected, we notice that the odds are typically between the true probabilities and the optimum probabilities.

Some interesting patterns emerge from \Cref{fig:2lineplot_prob}. The difference between the optimal probability of win and the probability of win with the actual team for Arsenal and Chelsea tend to be more as the season progresses. Although this indicates a progressively worse job in the team selection by the management, there are several other factors that may have contributed to this phenomena as well. The months of February and March pose extremely hectic schedules for top-ranked Premier League clubs who ply their trades in three other domestic and European championships along with the league. Naturally, it increases the chances of injury and fatigue for the players. While our data includes the information on suspension, long-term injury and subsequent non-availability of players, other instances of short-term absences are not recorded. 

For Aston Villa, Everton and Tottenham, we notice that the gaps in the two probabilities are small and uniform across the games. Only in the first few matches of the season, there are considerable differences, but the team managements did great jobs afterwards. It shows that while these teams were not among the strongest in that season (the optimal probabilities are not substantially high on all occasions), the management did a commendable job of picking their squad consistently. Contrast this with the graph for the two Manchester clubs and even more for Liverpool, where the differences were perpetually high. Historically, these teams are known to be strong, but during this particular season, their performances in the league were below the expectations. Our approach points out that a potential reason could be that the management of these clubs consistently selected far worse than optimal squads. The same is true for Sunderland even if they did not have a skilful squad, as the probability of wins are low in most matches.

The above analysis renders an excellent opportunity to quantify management efficiency of different clubs. The ratio of true probability to the optimum probability of winning is an appropriate yardstick in this aspect. It is always between 0 and 1, where a ratio closer to 1 (respectively 0) indicates good (respectively bad) decisions by the management. A summary of this measure for each team is provided via \Cref{tab:ratio_summary}. The teams are listed in decreasing order of average ratio and confirms the observations made earlier. Sunderland has the worst position in the list, with a low average score of management efficiency. A low average along with relatively high standard deviation of the ratio is noted for Stoke City, thereby suggesting not only weak management efficiency in terms of being able to field the best possible squads on average, but also a great deal of inconsistency in that domain.

 \begin{table}[!htb]
 \centering
 \caption{Summary of the management efficiency (calculated as the ratio of the originally selected team's probability  to the optimum probability) in the 2015/16 season for the ten clubs considered in this study. Teams are arranged in decreasing order of average efficiency.}
 \label{tab:ratio_summary}
 \begin{tabular}{lccccc}
   \hline
   Team & Minimum & Maximum & Average & Standard deviation & Median \\ 
   \hline
   Tottenham Hotspur & 0.319 & 0.970 & 0.822 & 0.124 & 0.862 \\ 
   Aston Villa & 0.243 & 0.920 & 0.746 & 0.136 & 0.771 \\ 
   Everton & 0.453 & 0.905 & 0.730 & 0.120 & 0.757 \\ 
   Arsenal & 0.429 & 0.983 & 0.722 & 0.111 & 0.725 \\ 
   Chelsea & 0.143 & 1.000 & 0.662 & 0.233 & 0.713 \\ 
   Manchester City & 0.258 & 0.915 & 0.647 & 0.141 & 0.656 \\ 
   Liverpool & 0.420 & 0.886 & 0.604 & 0.096 & 0.595 \\ 
   Manchester United & 0.153 & 0.983 & 0.579 & 0.230 & 0.591 \\ 
   Stoke City & 0.026 & 1.000 & 0.456 & 0.312 & 0.314 \\ 
   Sunderland & 0.092 & 0.798 & 0.403 & 0.160 & 0.400 \\ 
    \hline
 \end{tabular}
 \end{table}

Three big clubs -- Manchester United, Liverpool and Manchester City -- display comparable average results in this metric, placing them in the middle of the table. Arsenal and Everton are marginally better, and have higher management efficiency on average. The standard deviations are low for them, indicating consistent performances across the test set. Contrarily, Chelsea management perform similarly in terms of median but have more variation. Finally, Tottenham are found to record the best results in this regard. They have maintained excellent management efficiency across the season. Meanwhile, although the maximum probabilities of win for Aston Villa are on the lower half, the team can boast of the second most efficient management in selecting the optimal squad.

\subsection{A case study} \label{subsec:game:ex}

\begin{figure}[!htb]
     \caption{Match on 8th November 2015: Optimal teams for Tottenham Hotspur against Arsenal (left side) and for Arsenal against Tottenham Hotspur (right side).}
     \label{fig:tot_ars1}
     \centering
     \includegraphics[width = 14cm,keepaspectratio]{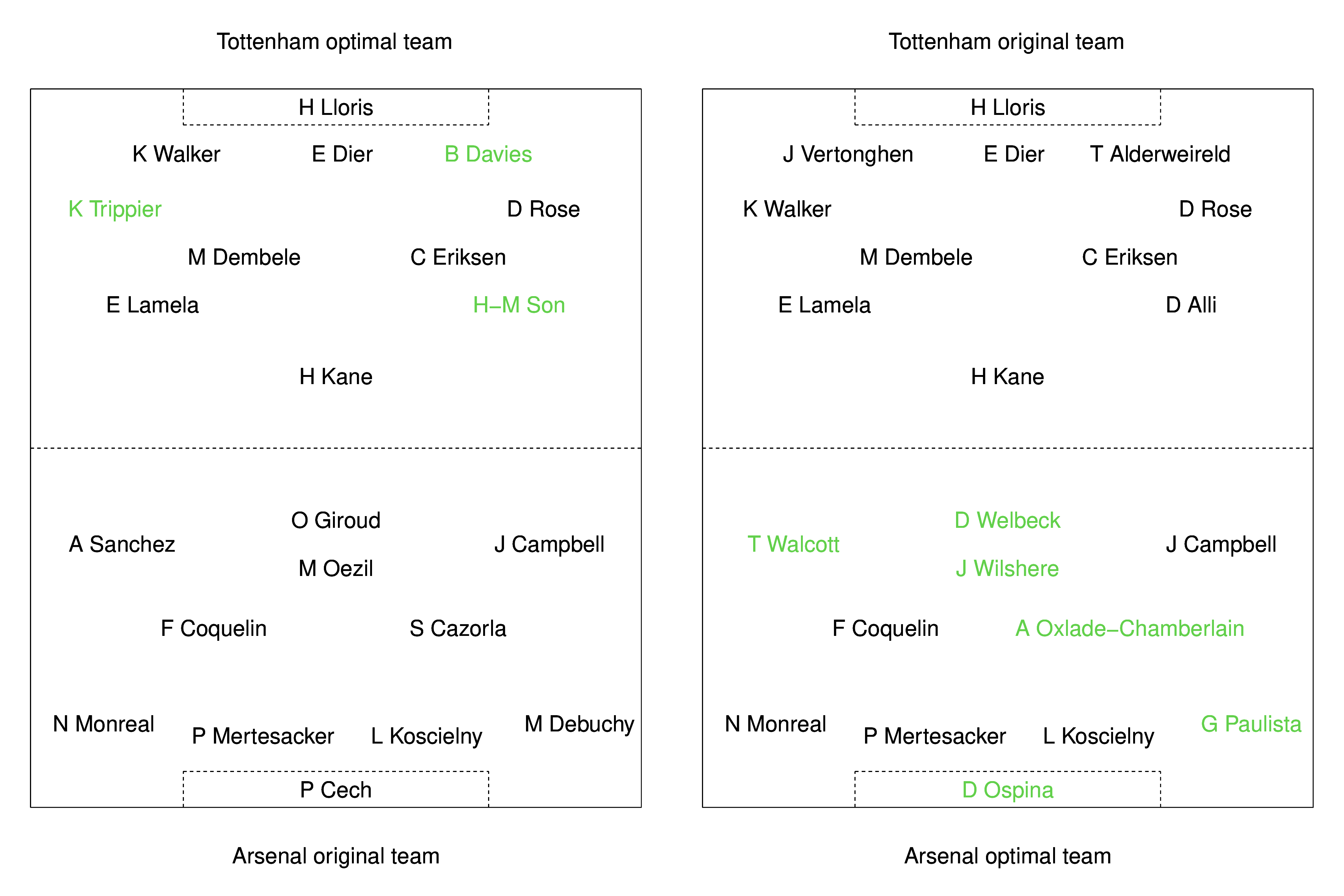}
 \end{figure}

\begin{figure}[!htb]
     \caption{Match on 5th March 2016: Optimal teams for Arsenal against Tottenham Hotspur (left side) and for Tottenham Hotspur against Arsenal (right side).}
     \label{fig:tot_ars2}
     \centering
     \includegraphics[width = 14cm,keepaspectratio]{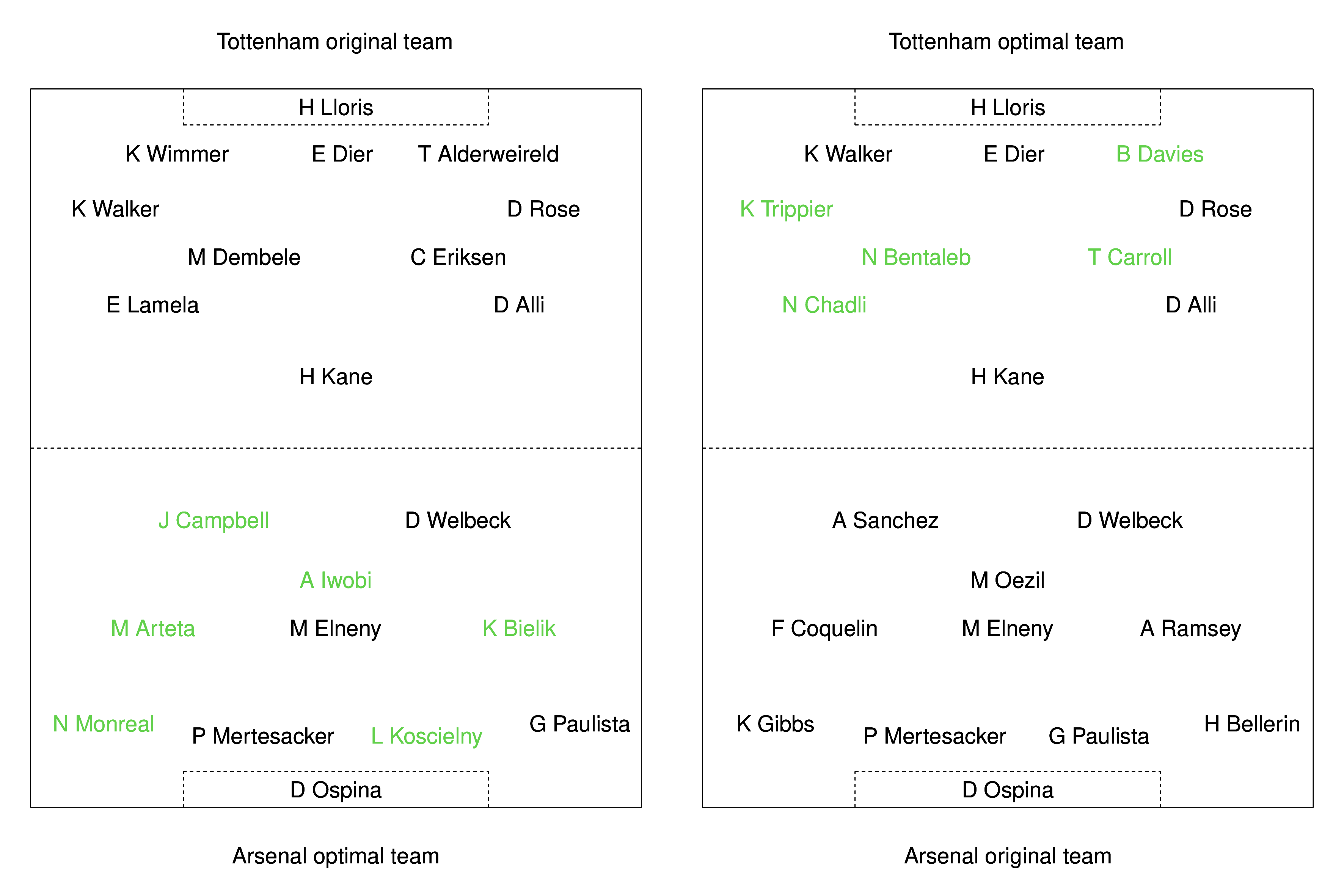}
 \end{figure}

Let us focus on an interesting match-up from the test set to demonstrate the efficacy of the proposed method. We consider the rivalry between two north London clubs, Arsenal and Tottenham, both of whom finished in the top table of 2015/16 season. Considering the home-and-away system in the league, we look at both matches between the two teams -- 
Arsenal's home match on 8th November 2015 ended with scoreline  1-1, while the reverse leg match held on 5th March 2016 had  scoreline 2-2. The original lineups of the two teams in these two matches, along with the optimal teams assuming the best-rated opponent lineups, are visible in Figures \ref{fig:tot_ars1} and \ref{fig:tot_ars2}.

In the first match, Arsenal played in a 4-3-3 formation. Our algorithm suggests six changes to the original lineup, which would increase its win probability to 73.9\% from 66.2\%. In midfield, Alex Oxlade-Chamberlain and Jack Wilshere should have been slotted in over Santi Cazorla and Mesut Oezil, while Danny Welbeck and Theo Walcott were more deserving to start up-front. Gabriel Paulista and David Ospina are the other two changes suggested by the method. Interestingly, we see that in the reverse leg, these two players were indeed included in the starting lineup. The team management, however, decided to drop Laurent Koscielny and Nacho Monreal, and that was not a good decision according to our method. In attack, Joel Campbell should have been included as well. At the centre of the park too, the Arsenal management should have chosen Alex Iwobi and youngster Krystian Bielik over out-of-form Mesut Oezil and Aaron Ramsey. Finally, Francis Coquelin, although was a good starter in the first leg, because of a dip in performance, should have been replaced by Mikel Arteta. Had these changes be included, Arsenal's chance of win could increase from 36.7\% to 57.1\% in its away game. 

We turn our attention to Tottenham lineups in the two matches. The management made only one change in the eleven, even though the matches took place five months apart, and that was not appropriate. In both games, Ben Davies and Kieran Trippier should have been included in the defending setup. Intriguingly, albeit Moussa Dembele and Christian Eriksen were correct choices in the first match, Nabil Bentaleb and Thomas Carroll would have been better picks in March. Contrastingly, Dele Alli needed to be replaced by Heung-Min Son in the first leg, but was the right choice later. We found that Tottenham's win probability in these two matches would rise to 36.8\% and 50.7\% from 27\% and 43.9\% respectively, had they incorporated these changes in the starting lineup.

\section{Conclusions} 
\label{sec:conclusion} 

Our motivation in this paper was to develop a unified framework of selecting lineup of a football team  that would serve two purposes. First, through a statistically appropriate model, we identify important features that contribute significantly to a team's performance. Alongside, the model also identifies pivotal individuals as well as some partnerships which impact a match's outcome, either positively or adversely. We use the data of most appearing EPL teams to demonstrate the effectiveness of the model. Although in the second stage of our algorithm we focus only on the probability of win, our findings in \Cref{subsec:model_summary} establish that treating the response variable as multinomial (with three categories as win, draw and loss) instead of binomial (win and not win) helps in deriving greater insight about the effects of various factors. The second important contribution of this research is to furnish a flexible algorithm that utilises the above model to arrive at the optimal starting lineup, depending on any possible opponent. Extensive exploration in \Cref{subsec:results_optimisation} reveals how the proposed algorithm can be used in practice. As a further by-product of the method, we have devised a measure to quantify management efficiency of a team in terms of its ability to select the optimal lineup. Similar techniques can be leveraged to find out a coach's decision-making ability and ingenuity in team selection. It would also render a way to discern if the team management are biased towards any player.

The flexible nature of the proposed methodology can be useful in various ways. For example, throughout this article, we consider only four primary positions for a player. The method can be easily modified to accommodate more specific roles such as right-wing-back, left-centre-half etc., which would be helpful to derive optimal lineup under specific strategies. In an identical manner, one can adjust the procedure to devise tactics for assigning optimal positions within a single game for players who are employable in multiple positions. Furthermore, the two-player interaction effect considered in this work can easily be extended to multiple-player combination effect, or to account for the interaction effect between two or more opposing players. Albeit that would lead to a larger set of variables, the computational burden is not expected to increase because of the LASSO step in the first stage. We also note that the substitutions of players are not considered in this work, but the current approach can surely lead to recommendations for good substitutions as well. 

Another interesting modification can be on the choice of the objective function. In this work, we have considered maximisation of win probability for the reference team. Alternatively, a lower-ranked team can be interested in minimising the chance of a loss against a higher-ranked club, whereas some team's modus operandi can be to maximise the expected number of points, i.e.\ the sum of the probability of drawing the game and three times the probability of winning the game. Our proposed methodology would work for such optimisation criterion too. Furthermore, under certain circumstances, a team management may deem some players to be essential in its starting lineup. Minor tweaking in our methodology will enable such framework where such players are always selected in the optimal lineup.

We now shift the focus from application to methodology and conclude with a brief account of possible modifications to the optimisation stage. Recall that the random starts in the algorithm follow the principles of stratified random sampling to find feasible solutions. Instead, random sampling with probability proportional to strength variables is an alternative possibility that may help with faster convergence. The algorithm can also be improved by considering a greedy solution at the beginning, where the best players from every position would be selected in the lineup. Determination of the best players can be according to the skills, the performances or other user-defined criteria. We have not considered either of these variations since the convergence was reasonably fast in our example, but they are expected to boost the speed further. 

The examples discussed in \Cref{subsec:game:ex} suggest other exciting avenues of future research too. In our implementation, we find out the optimal lineup of a team against another assuming that the latter plays its best-skilled players. This can be executed recursively to find out the optimal eleven for team A under the setting where team B is also expected to field its optimal eleven. It can be connected to the concepts of minimax risk, and is a fascinating direction to pursue further research. Another option is to select the two lineups from the perspectives of economic game theory. Last but not the least, the present work can be extended to other sports and we plan to pursue that in cases where it would require substantive and compelling modifications.

\section*{Data availability statement}

All data are extracted from the European soccer database (ESD), publicly available in Kaggle (link: \url{https://www.kaggle.com/hugomathien/soccer}).

\bibliography{References}

\newpage

\section*{Supplementary material}

\setcounter{table}{0}
\setcounter{section}{0}
\renewcommand{\thetable}{S\arabic{table}}
\renewcommand{\thesection}{S\arabic{section}}

\section{Comparison of different models}
\label{sec:results_comparison}

In this section, we compare the AIC values for the models discussed in the previous subsection to that of two other approaches that have been used in literature in assessing the effects of different variables on the multi-class outcomes in a soccer match. These values are reported in \Cref{tab:comparison-aic} below.

\begin{table}[!ht]
\centering
\caption{AIC of three competing models corresponding to the training data for different teams in the dataset.}
\label{tab:comparison-aic}
\begin{tabular}{l|c|c|c}
  \hline
  Team & LR model & PM rating based model & Proposed model \\ 
  \hline
  Arsenal & 486.017 & 490.890 & 475.688 \\ 
  Aston Villa & 500.740 & 507.365 & 499.090 \\ 
  Chelsea & 472.745 & 480.410 & 447.939 \\ 
  Everton & 563.376 & 544.211 & 541.283 \\ 
  Liverpool & 531.721 & 530.243 & 504.184 \\ 
  Manchester City & 448.686 & 462.126 & 436.286 \\ 
  Manchester United & 452.529 & 437.143 & 417.936 \\ 
  Stoke City & 528.014 & 516.428 & 469.054 \\ 
  Sunderland & 529.557 & 506.225 & 518.757 \\ 
  Tottenham Hotspur & 508.580 & 528.672 & 485.826 \\ 
   \hline
\end{tabular}
\end{table}

The first model in this comparative study is the multinomial logistic regression model with the rating variables, and without the proposed LASSO step in our method. \cite{carpita2019exploring} analysed the same European Soccer Database as ours and showed that such a regression model is usually better than other statistical or machine learning approaches. In the discussions below, we use the abbreviation LR to indicate this model. Next, in an attempt to understand whether our proposed skill variables are apt, we consider a different rating variable that has been proved to be effective in identifying good and bad players. It is the plus-minus (PM) rating variable that has been used in various sports, such as basketball, ice-hockey, soccer etc. In literature, there are several approaches to compute such rating variables. We measure it using a conventional way where the ratings are based on the goal differentials in presence of a particular player in the lineup (plus effect) and the same in absence of that player in the lineup (minus effect). It is noted that there are other methodologies to compute the plus-minus variables in soccer (see \cite{schultze2018weighted}, \cite{kharrat2020plus} for example). However, the main results are not expected to change substantially for those modifications.

The modelling strategy for assessing the effect of the PM rating variables is identical in nature to our proposed method. It is done to understand whether the PM ratings are better suited for our dataset than our skill variables and the player-wise fixed effects. In the first step if this approach, for every team and every match, the average PM ratings for every position are computed. Then, these variables and two-player interaction effects are considered in the LASSO framework as ours, where we force the model to include the PM ratings. Subsequently, after choosing the most appropriate set of features, a multinomial logistic regression model is run.

We compute the AIC values for the aforementioned two models and compare against the AIC of our approach. Recall that the AIC is a function of log-likelihood and number of parameters, and a lower AIC is desired. Now, from \Cref{tab:comparison-aic}, it is evident that the proposed model provides better fits in all cases, except for the data of Sunderland. We also note that the simple LR model is better than the PM rating based model in five cases whereas the latter beats the LR approach in the rest. Meanwhile, the LR model is outperformed by the proposed technique for all teams. 

To complete this comparison study, we focused on the scenario where the models are run on the combined dataset of all teams taken together. Looking at the performance of the three models, we found that the total AIC for the LR is around 4924.34, which is very close to the value for the PM rating based model too. In contrast, our model rendered an AIC value of 4678.97. Not only it confirms that the proposed method is better than other state-of-the-art techniques, but it also shows that individual team-wise analysis are significantly more suitable for some teams. In fact, it is natural to consider that the individual player effects and the synergy between certain players are specific to individual teams. Further, team-wise analysis do not consider the effects of the regressors to be universally the same across all the teams. This belief can be substantiated by looking at the results presented in the tables in the following section, where we can notice that the effect sizes of different variables are indeed starkly different for the ten teams.

Overall, the above analysis gives a fair idea about the effectiveness of our model in terms of the goodness of fit. One may thus conclude that the proposed rating variables as well as the LASSO-based multinomial logistic regression approach are suitable to analyse this dataset.

\section{Additional tables}

\Cref{tab:pca-weight} shows the weights as found by PCA-based approach for computing average skills in different categories.

\begin{table}[!hbt]
\centering
\caption{PCA-based weights for computing the average skills in different types of skills.}
\label{tab:pca-weight}
\begin{tabular}{lcc}
  \hline
    Type of skill & Attribute & Weight \\ 
    \hline
    Goalkeeping skill & diving & 0.207 \\ 
   & handling & 0.205 \\ 
   & kicking & 0.178 \\ 
   & positioning & 0.205 \\ 
   & reflexes & 0.204 \\ 
   \hline
  Defensive skill & interceptions & 0.237 \\ 
   & marking & 0.254 \\ 
   & standing tackle & 0.256 \\ 
   & sliding tackle & 0.253 \\ 
   \hline
  General skill & short passing & 0.091 \\ 
   & long passing & 0.070 \\ 
   & ball control & 0.105 \\ 
   & acceleration & 0.084 \\ 
   & sprint speed & 0.077 \\ 
   & agility & 0.093 \\ 
   & reactions & 0.076 \\ 
   & balance & 0.074 \\ 
   & jumping & 0.014 \\ 
   & stamina & 0.057 \\ 
   & strength & -0.026 \\ 
   & aggression & -0.000 \\ 
   & positioning & 0.091 \\ 
   & vision & 0.093 \\ 
   & dribbling & 0.103 \\ 
   \hline
  Attacking skill & crossing & 0.105 \\ 
   & finishing & 0.127 \\ 
   & heading accuracy & 0.003 \\ 
   & volleys & 0.132 \\ 
   & curve & 0.127 \\ 
   & free-kick accuracy & 0.123 \\ 
   & shot power & 0.124 \\ 
   & long shots & 0.139 \\ 
   & penalties & 0.120 \\ 
   \hline
\end{tabular}
\end{table}

Given below (in tables \ref{tab:arsenal} to \ref{tab:tottenham}) are details of the models selected for the ten teams based on model period. Since we predict recursively, these get updated, fine-tuned with additional match-data. 

\begin{table}[!htb]
\centering
\caption{Coefficient estimates of the fitted model for Arsenal} 
\label{tab:arsenal}
\begin{tabular}{lll}
  \hline
Variable & Win & Loss \\ 
  \hline
(Intercept) & $22.65 (0.003) ^*$ & $9.18 (0.004) ^*$ \\ 
  home & $0.42 (0.331) $ & $-0.75 (0.424) $ \\ 
  gk\_strength & $0 (0.046) $ & $0.03 (0.056) $ \\ 
  def\_strength\_def & $-0.01 (0.082) $ & $-0.2 (0.1) ^*$ \\ 
  mid\_strength\_att & $-0.12 (0.057) ^*$ & $-0.16 (0.07) ^*$ \\ 
  mid\_strength\_gen & $0.02 (0.087) $ & $0.06 (0.108) $ \\ 
  fwd\_strength\_att & $-0.02 (0.032) $ & $0.04 (0.04) $ \\ 
  gk\_strength\_opp & $0.03 (0.049) $ & $0.04 (0.062) $ \\ 
  def\_strength\_def\_opp & $-0.06 (0.058) $ & $-0.06 (0.071) $ \\ 
  mid\_strength\_gen\_opp & $-0.1 (0.084) $ & $0.1 (0.105) $ \\ 
  fwd\_strength\_att\_opp & $-0.04 (0.035) $ & $0.05 (0.042) $ \\ 
  Bacary Sagna(D):Denilson(M) & $-0.66 (0.443) $ & $-1.02 (0.56) $ \\ 
  Robin van Persie(F):Samir Nasri(M) & $-0.78 (0.531) $ & $-1.28 (0.684) $ \\ 
   \hline
\end{tabular}
\end{table}
\begin{table}[!htb]
\centering
\caption{Coefficient estimates of the fitted model for Aston Villa} 
\begin{tabular}{lll}
  \hline
Variable & Win & Loss \\ 
  \hline
(Intercept) & $-11.37 (0.021) ^*$ & $-12.68 (0.021) ^*$ \\ 
  gk\_strength & $-0.09 (0.079) $ & $-0.11 (0.076) $ \\ 
  def\_strength\_def & $0.17 (0.081) ^*$ & $0 (0.077) $ \\ 
  mid\_strength\_gen & $0.16 (0.089) $ & $0.1 (0.081) $ \\ 
  fwd\_strength\_att & $0.04 (0.066) $ & $0.07 (0.061) $ \\ 
  gk\_strength\_opp & $0.03 (0.047) $ & $0.08 (0.049) $ \\ 
  def\_strength\_def\_opp & $-0.05 (0.062) $ & $0.03 (0.06) $ \\ 
  mid\_strength\_def\_opp & $0.01 (0.026) $ & $-0.04 (0.024) $ \\ 
  mid\_strength\_gen\_opp & $-0.14 (0.088) $ & $-0.01 (0.085) $ \\ 
  fwd\_strength\_att\_opp & $0.05 (0.042) $ & $0.08 (0.039) ^*$ \\ 
  fwd\_strength\_def\_opp & $-0.04 (0.024) $ & $-0.06 (0.023) ^*$ \\ 
  Nathan Baker(D) & $0.6 (0.616) $ & $0.7 (0.577) $ \\ 
  Nathan Baker(D):Fabian Delph(M) & $0.57 (0.754) $ & $0.58 (0.706) $ \\ 
  Richard Dunne(D):Emile Heskey(F) & $-0.04 (0.801) $ & $-0.69 (0.82) $ \\ 
  James Collins(D):Emile Heskey(F) & $-1.03 (0.837) $ & $-0.16 (0.808) $ \\ 
   \hline
\end{tabular}
\end{table}
\begin{table}[!htb]
\centering
\caption{Coefficient estimates of the fitted model for Chelsea} 
\begin{tabular}{lll}
  \hline
Variable & Win & Loss \\ 
  \hline
(Intercept) & $-14.09 (0.006) ^*$ & $-7.64 (0.007) ^*$ \\ 
  home & $0.74 (0.355) ^*$ & $-0.91 (0.467) $ \\ 
  gk\_strength & $0 (0.068) $ & $-0.02 (0.081) $ \\ 
  def\_strength\_def & $0.06 (0.102) $ & $-0.14 (0.122) $ \\ 
  mid\_strength\_gen & $0.23 (0.079) ^*$ & $0.08 (0.094) $ \\ 
  fwd\_strength\_att & $0.05 (0.057) $ & $0.1 (0.071) $ \\ 
  gk\_strength\_opp & $0.02 (0.048) $ & $0.04 (0.062) $ \\ 
  def\_strength\_def\_opp & $0.04 (0.068) $ & $0.1 (0.08) $ \\ 
  mid\_strength\_gen\_opp & $-0.18 (0.084) ^*$ & $-0.03 (0.103) $ \\ 
  fwd\_strength\_att\_opp & $-0.04 (0.038) $ & $-0.01 (0.047) $ \\ 
  Florent Malouda(M) & $-1.33 (0.55) ^*$ & $-1.27 (0.662) $ \\ 
  Ashley Cole(D):Juan Mata(M) & $1.67 (0.677) ^*$ & $1.03 (0.802) $ \\ 
  Cesar Azpilicueta(D):Ramires(M) & $-0.39 (0.431) $ & $-1.66 (0.648) ^*$ \\ 
   \hline
\end{tabular}
\end{table}
\begin{table}[!htb]
\centering
\caption{Coefficient estimates of the fitted model for Everton} 
\begin{tabular}{lll}
  \hline
Variable & Win & Loss \\ 
  \hline
(Intercept) & $3.5 (0.007) ^*$ & $-4.1 (0.009) ^*$ \\ 
  home & $0.86 (0.326) ^*$ & $-0.48 (0.375) $ \\ 
  gk\_strength & $0.01 (0.085) $ & $0.02 (0.101) $ \\ 
  def\_strength\_def & $-0.03 (0.102) $ & $0.02 (0.113) $ \\ 
  mid\_strength\_gen & $0.05 (0.089) $ & $-0.01 (0.098) $ \\ 
  fwd\_strength\_att & $-0.03 (0.045) $ & $-0.05 (0.049) $ \\ 
  gk\_strength\_opp & $0.01 (0.042) $ & $-0.02 (0.047) $ \\ 
  def\_strength\_def\_opp & $-0.12 (0.06) ^*$ & $0.02 (0.065) $ \\ 
  mid\_strength\_def\_opp & $-0.02 (0.024) $ & $-0.07 (0.027) ^*$ \\ 
  mid\_strength\_gen\_opp & $0.13 (0.073) $ & $0.16 (0.084) $ \\ 
  fwd\_strength\_att\_opp & $-0.04 (0.035) $ & $-0.02 (0.038) $ \\ 
  Phil Jagielka(D):Sylvain Distin(D) & $-0.46 (0.354) $ & $-0.79 (0.399) ^*$ \\ 
  Leighton Baines(D):Marouane Fellaini(F) & $-0.56 (0.445) $ & $-0.45 (0.531) $ \\ 
  Tim Howard(G):Steven Pienaar(M) & $-0.87 (0.372) ^*$ & $-0.89 (0.411) ^*$ \\ 
  Phil Jagielka(D):Nikica Jelavic(F) & $0.16 (0.453) $ & $-1 (0.649) $ \\ 
   \hline
\end{tabular}
\end{table}
\begin{table}[!htb]
\centering
\caption{Coefficient estimates of the fitted model for Liverpool} 
\begin{tabular}{lll}
  \hline
Variable & Win & Loss \\ 
  \hline
(Intercept) & $4.04 (0.016) ^*$ & $-7.09 (0.021) ^*$ \\ 
  home & $-0.02 (0.332) $ & $-1.19 (0.404) ^*$ \\ 
  gk\_strength & $0.04 (0.049) $ & $0.09 (0.061) $ \\ 
  def\_strength\_def & $-0.08 (0.078) $ & $-0.06 (0.09) $ \\ 
  mid\_strength\_gen & $0.01 (0.079) $ & $0 (0.095) $ \\ 
  fwd\_strength\_att & $-0.02 (0.033) $ & $-0.06 (0.038) $ \\ 
  gk\_strength\_opp & $0.12 (0.048) ^*$ & $0.06 (0.054) $ \\ 
  def\_strength\_def\_opp & $-0.17 (0.068) ^*$ & $-0.13 (0.079) $ \\ 
  mid\_strength\_def\_opp & $-0.01 (0.023) $ & $-0.06 (0.027) ^*$ \\ 
  mid\_strength\_gen\_opp & $0.02 (0.073) $ & $0.25 (0.091) ^*$ \\ 
  fwd\_strength\_att\_opp & $0.04 (0.038) $ & $0.03 (0.044) $ \\ 
  Albert Riera(M) & $-1.29 (0.544) ^*$ & $-3 (0.917) ^*$ \\ 
  Fernando Torres(F):Dirk Kuyt(M) & $1.85 (0.674) ^*$ & $2.37 (0.762) ^*$ \\ 
   \hline
\end{tabular}
\end{table}
\begin{table}[!htb]
\centering
\caption{Coefficient estimates of the fitted model for Manchester City} 
\begin{tabular}{lll}
  \hline
Variable & Win & Loss \\ 
  \hline
(Intercept) & $40.77 (0.007) ^*$ & $48.23 (0.014) ^*$ \\ 
  gk\_strength & $-0.15 (0.104) $ & $-0.18 (0.116) $ \\ 
  def\_strength\_def & $-0.34 (0.108) ^*$ & $-0.49 (0.12) ^*$ \\ 
  mid\_strength\_gen & $0.13 (0.12) $ & $-0.01 (0.138) $ \\ 
  fwd\_strength\_att & $-0.04 (0.062) $ & $-0.09 (0.07) $ \\ 
  gk\_strength\_opp & $0.02 (0.051) $ & $0.05 (0.062) $ \\ 
  def\_strength\_def\_opp & $-0.2 (0.076) ^*$ & $-0.17 (0.087) $ \\ 
  mid\_strength\_gen\_opp & $0.07 (0.089) $ & $0.27 (0.11) ^*$ \\ 
  fwd\_strength\_att\_opp & $0.03 (0.038) $ & $0.02 (0.045) $ \\ 
  fwd\_strength\_def\_opp & $-0.01 (0.023) $ & $0.04 (0.026) $ \\ 
  Shaun Wright-Phillips(M) & $-1.24 (0.654) $ & $-2.26 (0.843) ^*$ \\ 
  Pablo Zabaleta(D):Yaya Toure(M) & $0.73 (0.486) $ & $0.41 (0.589) $ \\ 
  Joe Hart(G):Yaya Toure(M) & $0.67 (0.482) $ & $0.97 (0.608) $ \\ 
   \hline
\end{tabular}
\end{table}
\begin{table}[!htb]
\centering
\caption{Coefficient estimates of the fitted model for Manchester United} 
\label{tab:man-united}
\begin{tabular}{lll}
  \hline
Variable & Win & Loss \\ 
  \hline
(Intercept) & $10.71 (0.024) ^*$ & $-15.34 (0.038) ^*$ \\ 
  home & $2.03 (0.438) ^*$ & $0.71 (0.548) $ \\ 
  gk\_strength & $-0.09 (0.06) $ & $-0.08 (0.076) $ \\ 
  def\_strength\_def & $-0.01 (0.047) $ & $-0.05 (0.053) $ \\ 
  mid\_strength\_gen & $0.16 (0.094) $ & $0.34 (0.111) ^*$ \\ 
  fwd\_strength\_att & $-0.03 (0.044) $ & $-0.14 (0.055) ^*$ \\ 
  gk\_strength\_opp & $0.09 (0.054) $ & $0.06 (0.073) $ \\ 
  def\_strength\_def\_opp & $-0.12 (0.083) $ & $-0.04 (0.111) $ \\ 
  mid\_strength\_gen\_opp & $-0.14 (0.089) $ & $0.05 (0.12) $ \\ 
  fwd\_strength\_att\_opp & $0 (0.047) $ & $0.08 (0.059) $ \\ 
  Chris Smalling(D):Wayne Rooney(F) & $1.32 (0.719) $ & $0.95 (0.819) $ \\ 
  Chris Smalling(D):Michael Carrick(M) & $1.04 (0.722) $ & $-0.04 (0.862) $ \\ 
  Phil Jones(D):Michael Carrick(M) & $0.53 (0.651) $ & $1.01 (0.735) $ \\ 
  Rio Ferdinand(D):Rafael(D) & $0.51 (0.547) $ & $-0.81 (0.782) $ \\ 
  Nemanja Vidic(D):Darren Fletcher(M) & $0.5 (0.694) $ & $-0.86 (0.921) $ \\ 
  Patrice Evra(D):Darren Fletcher(M) & $0.09 (0.67) $ & $-0.8 (0.821) $ \\ 
  Rafael(D):Darren Fletcher(M) & $-0.24 (0.742) $ & $-2.19 (1.406) $ \\ 
  Jonny Evans(D):Antonio Valencia(M) & $0.3 (0.459) $ & $-1.28 (0.674) $ \\ 
  Michael Carrick(M):Juan Mata(M) & $-0.62 (1.035) $ & $0.49 (1.113) $ \\ 
   \hline
\end{tabular}
\end{table}
\begin{table}[!htb]
\centering
\caption{Coefficient estimates of the fitted model for Stoke City} 
\begin{tabular}{lll}
  \hline
Variable & Win & Loss \\ 
  \hline
(Intercept) & $54.25 (0.026) ^*$ & $51.14 (0.033) ^*$ \\ 
  home & $1.1 (0.396) ^*$ & $-1.11 (0.411) ^*$ \\ 
  gk\_strength & $-0.15 (0.089) $ & $-0.21 (0.092) ^*$ \\ 
  def\_strength\_def & $-0.15 (0.128) $ & $-0.15 (0.128) $ \\ 
  mid\_strength\_gen & $-0.32 (0.146) ^*$ & $-0.44 (0.142) ^*$ \\ 
  fwd\_strength\_att & $-0.04 (0.081) $ & $-0.05 (0.083) $ \\ 
  gk\_strength\_opp & $-0.02 (0.053) $ & $-0.04 (0.053) $ \\ 
  def\_strength\_def\_opp & $0.03 (0.068) $ & $0.21 (0.071) ^*$ \\ 
  mid\_strength\_gen\_opp & $-0.09 (0.082) $ & $0.09 (0.09) $ \\ 
  mid\_strength\_def\_opp & $-0.03 (0.027) $ & $-0.08 (0.029) ^*$ \\ 
  fwd\_strength\_att\_opp & $0.02 (0.042) $ & $-0.04 (0.042) $ \\ 
  Phil Bardsley(D) & $1.25 (0.791) $ & $1.3 (0.823) $ \\ 
  Mamady Sidibe(F) & $-2.24 (0.779) ^*$ & $-4.03 (0.839) ^*$ \\ 
  Robert Huth(D):Marc Wilson(D) & $-0.8 (0.562) $ & $-1.51 (0.553) ^*$ \\ 
  Andy Wilkinson(D):Peter Crouch(F) & $-0.27 (0.706) $ & $0.81 (0.735) $ \\ 
  Danny Higginbotham(D):Matthew Etherington(M) & $0.94 (0.673) $ & $-0.22 (0.776) $ \\ 
  Robert Huth(D):Matthew Etherington(M) & $0.65 (0.509) $ & $0.4 (0.499) $ \\ 
  Jonathan Walters(F):Glenn Whelan(M) & $0.15 (0.503) $ & $-2.34 (0.616) ^*$ \\ 
  Ricardo Fuller(F):Matthew Etherington(M) & $-0.56 (0.658) $ & $-2.41 (0.786) ^*$ \\ 
  Marko Arnautovic(M):Glenn Whelan(M) & $0.78 (0.609) $ & $-0.84 (0.696) $ \\ 
  Glenn Whelan(M):Rory Delap(M) & $-0.15 (0.577) $ & $2.01 (0.63) ^*$ \\ 
   \hline
\end{tabular}
\end{table}
\begin{table}[!htb]
\centering
\caption{Coefficient estimates of the fitted model for Sunderland} 
\begin{tabular}{lll}
  \hline
Variable & Win & Loss \\ 
  \hline
(Intercept) & $25.29 (0.014) ^*$ & $-7.75 (0.012) ^*$ \\ 
  home & $0.21 (0.368) $ & $-0.69 (0.338) ^*$ \\ 
  gk\_strength & $0.1 (0.074) $ & $0.13 (0.071) $ \\ 
  def\_strength\_def & $-0.14 (0.104) $ & $-0.15 (0.098) $ \\ 
  mid\_strength\_gen & $-0.2 (0.102) $ & $0.08 (0.093) $ \\ 
  fwd\_strength\_att & $0 (0.055) $ & $-0.05 (0.049) $ \\ 
  gk\_strength\_opp & $0.01 (0.057) $ & $0.05 (0.054) $ \\ 
  def\_strength\_def\_opp & $-0.01 (0.067) $ & $0.01 (0.061) $ \\ 
  mid\_strength\_gen\_opp & $-0.11 (0.081) $ & $0.06 (0.074) $ \\ 
  fwd\_strength\_att\_opp & $-0.01 (0.041) $ & $-0.02 (0.037) $ \\ 
  Phil Bardsley(D) & $0.86 (0.449) $ & $0.32 (0.409) $ \\ 
  Jack Colback(M) & $1.8 (0.662) ^*$ & $0.17 (0.68) $ \\ 
  Lee Cattermole(M) & $-0.04 (0.478) $ & $-0.63 (0.456) $ \\ 
  Phil Bardsley(D):Jack Colback(M) & $-0.9 (0.861) $ & $0.99 (0.86) $ \\ 
  Lee Cattermole(M):Sebastian Larsson(M) & $-0.64 (0.541) $ & $-0.55 (0.514) $ \\ 
  James McClean(M):Sebastian Larsson(M) & $-0.31 (0.554) $ & $-0.86 (0.528) $ \\ 
   \hline
\end{tabular}
\end{table}
\begin{table}[!htb]
\centering
\caption{Coefficient estimates of the fitted model for Tottenham Hotspur} 
\label{tab:tottenham}
\begin{tabular}{lll}
  \hline
Variable & Win & Loss \\ 
  \hline
(Intercept) & $25.9 (0.013) ^*$ & $34.65 (0.007) ^*$ \\ 
  home & $0.45 (0.364) $ & $-0.63 (0.417) $ \\ 
  gk\_strength & $0.19 (0.072) ^*$ & $-0.03 (0.072) $ \\ 
  def\_strength\_def & $-0.26 (0.117) ^*$ & $-0.06 (0.127) $ \\ 
  mid\_strength\_gen & $-0.04 (0.105) $ & $-0.25 (0.116) ^*$ \\ 
  fwd\_strength\_att & $-0.09 (0.084) $ & $-0.16 (0.091) $ \\ 
  gk\_strength\_opp & $-0.01 (0.058) $ & $0 (0.064) $ \\ 
  def\_strength\_def\_opp & $-0.19 (0.073) ^*$ & $-0.12 (0.081) $ \\ 
  mid\_strength\_gen\_opp & $0.11 (0.084) $ & $0.16 (0.099) $ \\ 
  fwd\_strength\_att\_opp & $-0.03 (0.04) $ & $0.03 (0.044) $ \\ 
  fwd\_strength\_def\_opp & $-0.02 (0.022) $ & $-0.07 (0.028) ^*$ \\ 
  Michael Dawson(D):Jermain Defoe(F) & $-0.55 (0.564) $ & $0.06 (0.631) $ \\ 
  Younes Kaboul(D):Aaron Lennon(M) & $1.36 (0.63) ^*$ & $1.73 (0.668) ^*$ \\ 
  Ledley King(D):Benoit Assou-Ekotto(D) & $1.86 (0.574) ^*$ & $1.48 (0.627) ^*$ \\ 
  Jermain Defoe(F):Heurelho Gomes(G) & $0.16 (0.665) $ & $-0.75 (0.75) $ \\ 
  Roman Pavlyuchenko(F):Heurelho Gomes(G) & $0.86 (0.551) $ & $-0.29 (0.702) $ \\ 
   \hline
\end{tabular}
\end{table}

\end{document}